\documentstyle[aps,prl]{revtex}
\def \d {{\rm d}}
\begin{document}
\draft
\title{Chaos in pp-wave spacetimes}
\author{Ji\v{r}\'{\i} Podolsk\'y,\  Karel Vesel\'y}
\address{Department of Theoretical Physics,
Faculty of Mathematics and Physics, Charles University,\\
V Hole\v{s}ovi\v{c}k\'ach 2, 180~00 Prague 8, Czech Republic.}

\date{\today}
\twocolumn[
\maketitle
\widetext
\begin{abstract}
We demonstrate chaotic behavior of timelike, null and spacelike
geodesics in non-homogeneous vacuum {\it pp}-wave solutions.
This seems to be the first known example of a chaotic motion
in exact radiative spacetime.
\end{abstract}

]
\narrowtext

In the context of general relativity the first system for which
a chaotic behavior of solutions to the Einstein equations has been
recognized and thoroughly studied were Bianchi IX
cosmological models (see for example
\cite{Hol}, \cite{CL1}, and references therein).
Complicated nonlinear effects also occur in systems with
coupled gravitational and scalar fields \cite{Chop} - \cite{CL}.

Another type of problems providing nonlinear dynamical systems
in general relativity are the studies of geodesic motion in given spacetimes.
In particular, chaotic behavior of geodesics in the relativistic analogue
of the two fixed-centres problem (modelled by extreme black holes)
was examined in\cite{Conto1} - \cite{CG}. Chaotic geodesic motion was also
found in Schwarzschild spacetime \cite{BoCa} - \cite{SuMa},
in some static axisymmetric spacetimes \cite{KaVo}, \cite{SSM}
and in topologically non-trivial Robertson-Walker universe
\cite{LMP}, \cite{Tom}.

Here we investigate motion in exact gravitational waves,
namely in the widely known class of vacuum {\it pp}-waves  \cite{KSMH},
the metric of which can be written in the form
\begin{equation}
\d s^2=2\,\d\zeta \d\bar\zeta-2\,\d u\d v-(f+\bar f)\,\d u^2\ , \label{E1}
\end{equation}
where  $f(u,\zeta)$ is an arbitrary function of $u$ and
the complex coordinate $\zeta$ spanning the plane wave surfaces $u=$ const.
When $f$ is linear in $\zeta$, the metric (\ref{E1}) represents
Minkowski universe. The case $f=g(u)\zeta^2$ describes
plane gravitational waves (`homogeneous' {\it pp}-waves) which
have thoroughly been investigated (see \cite{KSMH} for Refs).
This simple example of an exact radiative spacetime has also
been used for construction of sandwich and impulsive waves
\cite{BPR}, \cite{Penrose}.

However, here we wish to study geodesics in more general,
non-homogeneous vacuum {\it pp}-waves and demonstrate
their chaotic behavior. The geodesic equations for (\ref{E1}) are
\begin{eqnarray}
&&\ddot\zeta + \textstyle{\frac{1}{2}}\bar f_{,\bar\zeta}\, U^2 = 0 \ ,\label{E2}\\
&&\dot u=U=\hbox{const.} \ ,\label{E3}\\
&&\ddot v+(f_{,\zeta}\dot\zeta+\bar f_{,\bar\zeta}\dot{\bar\zeta})\, U
 +\textstyle{\frac{1}{2}}(f+\bar f)_{,u}\, U^2=0 \ ,\label{E4}
\end{eqnarray}
where dot denotes $d/d\tau$ with $\tau$ being an affine parameter.
Assuming also a condition normalizing
the four-velocity such that $U_\mu U^\mu=\epsilon$ where
$\epsilon=-1, 0, +1$ for timelike, null or spacelike geodesics,
we get
\begin{equation}
\dot v=\textstyle{\frac{1}{2}}U^{-1}\left[2\dot\zeta \dot{\bar\zeta}
  -(f+\bar f)\,U^2-\epsilon\right]\ . \label{E5}
\end{equation}
We consider $U\not=0$ (for $U=0$ Eqs. (\ref{E2})-(\ref{E4})
can be integrated yielding only some trivial geodesics).
By differentiating Eq. (\ref{E5})  and using (\ref{E2})
we immediately obtain (\ref{E4}) which can thus be omitted.
Hence it suffices to find solutions of Eq. (\ref{E2}) since  $v(\tau)$
can then be obtained by integrating Eq. (\ref{E5}), and $u(\tau)=U\tau+u_0$.

The remaining equation (\ref{E2}) has the same form for timelike,
null and spacelike geodesics. Introducing real coordinates $x$ and $y$
by $\zeta=x+\hbox{i}y$ we get a system which, for $f$ independent of $u$,
follows from the Hamiltonian
\begin{equation}
H=\textstyle{\frac{1}{2}}\left(p_x^2+p_y^2 \right)+V(x, y)\ , \label{E7}
\end{equation}
where the potential is $V(x,y)=\textstyle{\frac{1}{2}}U^2\, {\cal R}e\,f$.
For non-homogeneous {\it pp}-wave spacetimes given by
$f=C\zeta^n$, $C=\hbox{const.}>0$, $n=3, 4, \cdots$,
the corresponding potential
\begin{equation}
V(x, y) =\textstyle{\frac{1}{2}}CU^2\, {\cal R}e\,\zeta^n\ , \label{E7b}
\end{equation}
is called `$n$-saddle'. It can by visualized in polar coordinates
$\rho$, $\phi$ where $\zeta=\rho\exp(\hbox{i}\phi)$,
in which it takes the form
$V(\rho,\phi)=\textstyle{\frac{1}{2}}CU^2\, \rho^n\cos(n\phi)$.

Now, it was shown previously by Rod, Churchill and Pecelli
in a series of papers \cite{Rod} - \cite{CR1} that motion in
the Hamiltonian system (\ref{E7}) with the $n$-saddle potential
(\ref{E7b}) is chaotic.

Let us first briefly summarize their results for the simplest case
$n=3$. The corresponding potential (after removing
an unimportant multiplicative factor by a suitable rescaling of $\tau$)
\begin{equation}
V(x, y) =\textstyle{\frac{1}{3}}x^3-xy^2 \label{E8}
\end{equation}
is called a `monkey saddle'. Interestingly, this is a special case of famous
H\'enon-Heiles Hamiltonian \cite{HH} which is known to be a `textbook'
example of a chaotic system (however, their quadratic terms are
absent in our case). This particular case of the H\'enon-Heiles
Hamiltonian has been investigated by Rod \cite{Rod}.
He concentrated on bounded orbits in the energy
manifolds $H(x, y, p_x, p_y)=E>0$. The homogeneity of $V$
guarantees that the orbit structure for any two positive values of
$E$ is isomorphic modulo a constant scale factor and adjustment of time:
$x\to \tilde x= \lambda x$, $y\to \tilde y= \lambda y$ and
$\tau\to \tilde\tau= \tau/\sqrt\lambda$ results in
$E\to \tilde E= \lambda^3 E$. Therefore, without loss of generality
one can restrict to one particular value of $E$.

In order to describe the topological structure of all bounded orbits
Rod first constructed  three basic {\it unstable periodic orbits}
(denoted by $\Pi_j$) which are {\it isolated invariant sets} for the flow.
The region in which these bounded orbits occur can be decomposed into three
disjoint cells $R_j$  (see \cite{Rod} or \cite{PoVe} for details);
each contain only one orbit $\Pi_j$ and no other bounded orbits.
Hence, $\Pi_j$ is the only invariant set in $R_j$, i.e. it is isolated.

Subsequently, Rod investigated {\it orbits asymptotic to}
basic orbits $\Pi_j$ as $\tau\to\pm\infty$
and showed that these asymptotic sets intersect transversely.
This gives the existence of  orbits that
`connect' the orbits $\Pi_j$:
they are {\it homoclinic} (asymptotic to the same periodic orbit
in both time directions) or {\it heteroclinic} (asymptotic to
two different periodic orbits, one in each time direction).
It is the existence of these orbits that illustrates complicated chaotic
structure of the flow.

The topology of possible orbits in phase space can also be represented
by symbolic dynamics given here by a set of biinfinite sequences,
$s\equiv\cdots, s_k, s_{k+1}, s_{k+2}, \cdots$, where $s_k\in\{1,2,3\}$,
$s_k\not=s_{k+1}$. Using a topological version of the Smale horseshoe
map, it was shown in \cite{Rod} that to any biinfinite sequence of symbols
$\{1, 2, 3\}$ there exists an {\it uncountable number of bounded orbits}
running through the blocks $R_j$ in the prescribed order as $\tau$ goes from
$-\infty$ to $+\infty$ . Also, the flow admits at least a countable
number of non-degenerate homoclinic and heteroclinic orbits.

Rod remarked that these results could be refined if the unstable
periodic orbits $\Pi_j$ were known to be hyperbolic so that they would
admit stable and unstable asymptotic manifolds. Consequently,
to each periodic symbol sequence there would correspond a countable
collection of {\it periodic} orbits. The difficulties in proving
the hyperbolicity of $\Pi_j$ were subsequently overcome in
\cite{RPC}. In \cite{CR3}, summarizing and generalizing some previous results
\cite{CR1}, the Hamiltonian (\ref{E7}), (\ref{E8}) was presented as an example
of a system for which the Smale horseshoe map can {\it explicitly}
be embedded as a subsystem into the flow along the homoclinic
and heteroclinic orbits. This completed the proof of chaotic
behavior of the studied system.

Similar results hold for geodesic motion in arbitrary
non-homogeneous {\it pp}-waves with the structural function
of the form $f=C\zeta^n$, where $n\ge4$, i.e. for a general $n$-saddle
potential (\ref{E7b}). It was shown in \cite{Rod}, \cite{CR1}
that the decomposition into isolating cells $R_j$,
$j=1, 2, \cdots, n$, each containing exactly one of the basic
unstable periodic solutions $\Pi_j$, is analogous to the case
$n=3$. Subsequently, the orbits $\Pi_j$ were
proven to be hyperbolic \cite{RPC} and the existence of
homoclinic and heteroclinic orbits was established
\cite{CR3}. Again, given any biinfinite sequence, uncountably many
orbits can be found which pass from one  block containing $\Pi_j$
to the other in the specified order.

In order to independently support these arguments for chaotic
behavior of geodesics in non-homogeneous {\it pp}-waves we
investigate the motion also by a fractal method.
Complementary to the analysis described above, we concentrate
on {\it unbounded} geodesics and we do not restrict to the same
energy manifold $E=$ const.

Chaos is usually indicated by a sensitive dependence of the
evolution on the  choice of initial conditions.
The coordinate independent fractal method (see for example
\cite{CL1}, \cite{Conto1}, \cite{DFC}, \cite{CG}) starts with a
definition of different asymptotic outcomes (given here by
`types of ends' of all possible trajectories). A set of initial
conditions is evolved numerically until one of the outcome states is
reached. Chaos is uncovered if the basin boundaries which separate
initial conditions leading to different outcomes are fractal.
Such fractal partitions are the result of chaotic dynamics.
As we shall now demonstrate, we observe exactly these structures
in the studied system.

We integrate numerically the equations of motion given by
(\ref{E7}), (\ref{E7b}). The initial conditions are chosen such that
the geodesics start at $\tau=0$ from a unit circle in the
($x, y$)-plane (due to the homogeneity of the $n$-saddle potential
all other geodesics can simply be obtained by a suitable rescaling).
It is natural to parametrize the initial positions by an angle
$\phi\in [-\pi,\pi)$ such that $x(0)=\cos\phi$, $y(0)=\sin\phi$.
In Fig.~1 we present typical trajectories of geodesics for
$n=3, 4, 5$,  when $\dot x(0)=0=\dot y(0)$.
We observe that each unbounded geodesic escapes
to infinity (where for $n\ge3$ the curvature singularity is located)
{\it only along } one of the $n$ channels in the potential.
The axes of these outcome channels are given in polar coordinates
by the condition $\cos(n\phi_j)=-1$, $j=1, \cdots, n$, and represent
radial lines `of steepest descent' since $V\to-\infty$ as $\rho\to\infty$
most rapidly along them. (For non-zero initial velocities more
geodesics prefer one of the channels but the character of motion
does not change significantly \cite{PoVe}.)

In fact, any unbounded geodesic oscillates around the radial axis
$\phi_j=(2j-1)\pi/n$ of the corresponding $j$-th outcome channel.
Introducing $\Delta\phi_j(\tau)=\phi(\tau)-\phi_j$ we find
asymptotically that
$\rho\approx[(\frac{n}{2}-1)\sqrt{CU^2}(\tau_s-\tau)]^{2/(2-n)}$
as $\rho\to\infty$ and
\begin{eqnarray}
&&\Delta\phi_j(\tau)\approx (\tau_s-\tau)^\alpha
        \Big(A\cos\left[b\ln(\tau_s-\tau)  \right] \nonumber\\
&&\hskip29mm+B\sin\left[b\ln(\tau_s-\tau)\right]\Big)\ ,\label{E13}
\end{eqnarray}
where $\alpha=\frac{1}{2}(n+2)/(n-2)$, $b=\frac{1}{2}\sqrt{7n^2-4n-4}/(n-2)$
and $A$, $B$ are constants. As the geodesics approach
the singularity at $\rho=\infty$, $\tau\to\tau_s$, frequency of their
oscillations around $\phi_j$ grows to infinity while the
amplitude of  oscillations tends to zero. We may call this effect
a `focusation'.

\newpage

\
\vspace*{-10mm}
\begin{figure}[h]
\hspace{10mm}
\special{em:graph 98fig1.pcx}
\vspace{163mm}
\caption{ Geodesics starting from a unit circle escape to infinity
only along one of the $n$ channels.}
\end{figure}

Let us return back to our observation that all unbounded geodesics
approach infinity through only $n$ {\it distinct channels}. These represent
possible outcomes of our system and we assign them symbol
$j$ which takes one of the corresponding values,
$j\in\{1, 2, \dots, n\}$ (thus, for example, $j=1$ means that the
geodesic approach infinity at $\rho=\infty$ through the first
channel with the axis $\phi_1=\pi/n$ as $\tau\to\tau_s>0$).
From Fig.~1 we observe that in certain regions the function
$j(\phi)$ depends  very sensitively on initial position given by $\phi$.
We calculated $j(\phi)$ numerically for $n=3, 4, 5$ --- the results
are shown in Fig.~2. Also, in the same diagrams we plot
the function $\tau_s(\phi)$ which takes the value of the parameter
$\tau$ when the singularity at $\rho=\infty$ is reached by a given geodesic.

\
\vspace*{-10mm}
\begin{figure}[h]
\hspace{4mm}
\special{em:graph 98fig2.pcx}
\vspace{163mm}
\caption{The functions $j(\phi)$ and $\tau_s(\phi)$
indicate that basin boundaries separating different outcomes
are fractal.}
\end{figure}

Clearly, the boundaries between the outcomes appear to be fractal
which can be confirmed on the enlarged detail of the image and the
enlarged detail of the detail etc. In Fig.~3 we show such zooming in
of the fractal interval localized around the value $\phi\approx0$
for $n=3$ (there are two symmetric fractal intervals in this case
around $\phi\approx\pm\frac{2}{3}\pi$) up to the sixth level.
At {\it each level} the structure has the same property,
namely that between two larger connected sets of geodesics with
outcome channels $j_1$ and $j_2\not=j_1$ there is always
a smaller connected set of geodesics with outcome
channel $j_3$ such that $j_3\not=j_1$ and $j_3\not=j_2$.
Similarly as in \cite{Conto1}, \cite{CG}, the structure of initial conditions
resembles three mixed Cantor sets,
and this fact is a manifestation of chaos.

\
\vspace*{-6mm}
\begin{figure}[h]
\special{em:graph 98fig3.pcx}
\vspace{112mm}
\caption{The fractal structure described by $j(\phi)$ and $\tau_s(\phi)$
is clearly confirmed here by zooming in the interval around $\phi\approx0$
for $n=3$.}
\end{figure}

The above structure of $j(\phi)$ has its counterpart in
the fractal structure of $\tau_s(\phi)$, see Fig.~2 and Fig.~3.
We observe that the value of $\tau_s$ goes to infinity
on each discontinuity of $j(\phi)$, i.e., on any fractal basin boundary
between the different outcomes.
There is an infinite set of peaks corresponding to chaotic bounded
orbits which never `decide' on a particular outcome, and so never escape
to infinity. The value of $\tau_s$ also increases in non-chaotic regions
of $\phi$ as one zooms in the higher levels of the fractal.
This is natural since these levels are given by
geodesics which undergo `more bounces' on the potential walls
before falling into one of the outcome channels so that their values
of $\tau_s$ are greater.

We demonstrated chaotic behavior of geodesics
in non-homogeneous {\it pp}-waves by invariant
analytic and numerical methods. As far as we know, this is the
first explicit demonstration of chaos in {\it exact} radiative spacetimes
(chaotic interaction of particles with linearized gravitational waves
on given backgrounds has already been studied in
\cite{BoCa}, \cite{LeVi}, \cite{VaPa} - \cite{Koku}). Since {\it pp}-wave
solutions are the simplest gravitational waves it would be an
interesting task to search for a chaotic motion in other radiative
space-times.

We acknowledge the support of grants GACR-202/
96/0206 and GAUK-230/1996.

\end{document}